\documentclass[10pt,conference]{IEEEtran}
\usepackage{xcolor,soul,framed} 
\usepackage[noadjust]{cite}
 
\colorlet{shadecolor}{yellow}
\usepackage[pdftex]{graphicx}
\graphicspath{{../pdf/}{../jpeg/}}
\DeclareGraphicsExtensions{.pdf,.jpeg,.png}
\usepackage[caption=false, font=footnotesize]{subfig}
\usepackage{amssymb,mathtools}
\usepackage{array}
\usepackage{mdwmath}
\usepackage{mdwtab}
\usepackage{eqparbox}
\usepackage{url}
\usepackage[ruled,vlined,linesnumbered]{algorithm2e}
\makeatletter
\let\OriginalAlgorithmTopRule\@algocf@pre@ruled
\renewcommand*{\@algocf@pre@ruled}{
\vspace*{5pt}
\OriginalAlgorithmTopRule
}
\makeatother
\usepackage{enumitem} 
\usepackage[mathcal]{euscript} 
\usepackage{tabularx} 
\usepackage{multirow}
\usepackage{booktabs}
\usepackage{makecell}
\usepackage{aligned-overset}
\usepackage{url}
\usepackage{multicol}
\usepackage{algorithmic}
\usepackage{balance}
\usepackage{bbm}
\usepackage{amsmath}
\usepackage{amsthm}
\usepackage{pifont}

\newtheoremstyle{mytheorem}
  {0pt}
  {0pt}
  {\normalfont}
  {}
  {\bfseries\itshape}
  {.}
  {0.5em}
  {}
\theoremstyle{mytheorem}
\graphicspath{{./Figures/}}

\usepackage[hidelinks]{hyperref}
\usepackage{xcolor}

\usepackage[letterpaper,left=0.635in,right=0.635in,top=0.75in,bottom=1.05in]{geometry}
\begin{document}
\title{User-Target Pairing for NOMA-Enabled ISAC \\ in Low-Altitude Wireless Networks}
\author{
  \IEEEauthorblockN{Irfan Azam, Muddasir Rahim, and Soumaya Cherkaoui}
  \IEEEauthorblockA{Department of Computer and Software Engineering, Polytechnique Montreal, Canada} \IEEEauthorblockA{Email: \{\texttt{irfan.azam, muddasir.rahim, soumaya.cherkaoui}\}@polymtl.ca}
}

\maketitle
\begin{abstract}
This paper presents a non-orthogonal multiple access (NOMA)-enabled integrated sensing and communication (ISAC) framework for low-altitude wireless networks (LAWN), where an aerial access point simultaneously serves ground users and senses aerial targets. To improve both communication efficiency and sensing reliability, a matching-based user-target pairing (MUTP) scheme is proposed, in which strong and weak channel NOMA users are first paired into NOMA groups and then assigned to sensing targets according to a sensing-aware pairing criterion. Furthermore, an adaptive power allocation (APA) strategy is developed to balance communication quality-of-service requirements and target detection performance. The proposed MUTP-APA scheme is compared with fixed greedy pairing, exhaustive search pairing, and conventional ISAC benchmark schemes under both fixed power allocation and APA. Simulation results show that our proposed scheme achieves higher sum rate, improved worst-target detection probability, and lower communication and sensing outage probabilities compared with conventional ISAC and fixed pairing baselines in NOMA-ISAC LAWN. Moreover, the proposed method closely approaches the exhaustive search benchmark while avoiding the high computational complexity.
\end{abstract}
\begin{IEEEkeywords}
integrated sensing and communication, low-altitude wireless networks, non-orthogonal multiple access, user-target pairing.
\end{IEEEkeywords}
\section{Introduction}
The rapid development of sixth-generation (6G) wireless networks and the low-altitude economy is driving the need for wireless infrastructures that can simultaneously support connectivity, sensing, computation, control, and airspace awareness. Low-altitude wireless networks (LAWNs) have recently emerged as a promising architecture for enabling intelligent services in three-dimensional (3D) low-altitude airspace, where unmanned aerial vehicles (UAVs), electric vertical take-off and landing aircraft, aerial robots, and low-altitude platforms (LAPS) require reliable communication and real-time environmental perception \cite{Wu_LAWN_Survey,Jin_LAWN_SC2}. Unlike conventional terrestrial networks, LAWNs must cope with dynamic user and target distributions, mobility-induced channel variations, line-of-sight (LoS)-dominant propagation, and strict reliability requirements. These characteristics make communication-only network designs insufficient for supporting future LAWN applications.

Integrated sensing and communication (ISAC) is a key enabling technology for LAWNs because it allows wireless communication and radar sensing to share spectrum, hardware, and waveform resources \cite{Liu_ISAC_Survey,Liu_ISAC_JSAC,Cui_ISAC_IoT}. In an ISAC-enabled LAWN, the access point (AP) or aerial platform can simultaneously serve communication users (CUs) and sense surrounding targets for localization, tracking, and airspace monitoring. Recent studies have investigated LAWN-oriented ISAC designs from different perspectives, including sensing-communication-control, full-duplex integrated sensing, communication, and computation over LAPS~\cite{Chen_FD_ISCC_LAWN}, and networked ISAC for low-altitude economy oriented transmit beamforming and UAV trajectory optimization \cite{Cheng_Networked_ISAC_LAE}. These works demonstrate that sensing and communication performance are strongly coupled in LAWN scenarios and must be jointly optimized rather than treated as independent functions.

Although ISAC improves spectrum and hardware efficiency, supporting multiple CUs and multiple sensing targets (STs) over limited wireless resources remains challenging. Orthogonal multiple access (OMA)-based schemes can reduce inter-user interference by assigning users to separate time/frequency resources, but they may suffer from low spectral efficiency when the number of CUs and STs increases. To cope with the challenges in OMA, one promising solution is non-orthogonal multiple access (NOMA), which allows multiple users to share the same resource block through power domain (PD) superposition and successive interference cancellation (SIC) \cite{Ding_NOMA,Zhang_Semi_ISAC_NOMA}. In the context of ISAC, NOMA can further enhance resource utilization because the superimposed communication waveform can also contribute to sensing \cite{Mu_NOMA_ISAC_Perspective,Ouyang_SIC_NOMA_ISAC,Wang_NOMA_Empowered_ISAC}. The performance of downlink and uplink NOMA-ISAC systems has been analyzed in terms of outage probability, ergodic rate, sensing rate, and target detection probability \cite{Ouyang_Downlink_NOMA_ISAC,Agrawal_NOMA_ISAC_Rician}. These studies confirmed the potential of NOMA for improving the communication-sensing tradeoff compared with orthogonal resource partitioning.

Despite these advantages, the performance of NOMA-ISAC critically depends on how CUs are paired and how STs are associated with the formed NOMA groups. Meanwhile, in conventional NOMA systems, CUs are often paired according to their channel disparity, where a near CU with a strong channel is paired with a far CU with a weak channel to facilitate SIC and improve fairness \cite{Zhu_NOMA_User_Pairing}. Therefore, this communication-oriented rule is not sufficient for ISAC-based LAWNs. A NOMA pair that is favorable for CU may generate a beam that is poorly aligned with a ST, leading to weak target illumination and degraded detection probability. Conversely, assigning targets only according to sensing suitability may degrade the communication rate or violate QoS constraints. Hence, LAWN-based NOMA-ISAC requires a joint design that accounts for both CU pairing and ST assignment.

Recent NOMA-aided ISAC works have started to investigate CU-ST pairing, power control, and beamforming. For example, multi-armed bandit-based approaches have been proposed to jointly optimize CU-ST pairing and beamforming in NOMA-aided ISAC networks \cite{Nasser_MAB_UTP, Liu_UPPA, Nasser_UTP_TCCN}. Moreover, authors in~\cite{Dong_CFMM_NOMA_ISAC} studied joint user pairing and beamforming for NOMA-aided cell-free massive multiple-input multiple-output ISAC systems. These studies highlight the importance of pairing in NOMA-ISAC, but they mainly focus on generic terrestrial or cell-free architectures and often rely on learning-based or iterative optimization methods. However, LAWN specific aspects, such as low-altitude mobility, target distance dependent sensing difficulty, and scalable pairing among multiple CUs and multiple STs, remain underexplored.

\subsection{Motivation and Contributions}

Motivated by the coupled communication-sensing requirements of LAWNs, this paper investigates joint user pairing, target assignment, and power allocation for NOMA-ISAC systems. In NOMA-ISAC, communication performance depends on strong-weak CU pairing, while sensing reliability depends on the suitability of assigning STs to NOMA groups. Fixed GP schemes are simple but cannot adapt to this coupling, whereas exhaustive search provides an upper-bound solution at rapidly increasing complexity. Therefore, we propose an MUTP-APA framework to improve communication efficiency and sensing reliability without exhaustive enumeration.

The main contributions of this paper are as follows.
\begin{itemize}
    \item We develop a NOMA-ISAC system model for LAWNs and formulate the joint user-pairing, target-assignment, and power allocation problem under 3D distance-dependent propagation, SIC feasibility, communication QoS, and target detection constraints.

    \item We propose an MUTP scheme that forms communication-aware strong-weak NOMA CU pairs using QoS-aware utilities and assigns STs to NOMA groups using a sensing-aware max-min association criterion.

    \item We develop a QoS and sensing-aware APA strategy that evaluates multiple candidate power profiles, including fixed, sensing-aware, QoS-aware, and communication-oriented allocations, and selects the best profile based on sum rate, QoS satisfaction, and sensing reliability.

    \item We evaluate the proposed NISAC-MUTP-APA scheme against GP, exhaustive-search, and conventional ISAC benchmarks. Simulation results show improved sum rate, worst-target detection probability, communication outage, and sensing outage, while closely approaching exhaustive-search performance with lower complexity.
\end{itemize}

\begin{figure}[!t]
     \centering
\includegraphics[width=0.85\linewidth]{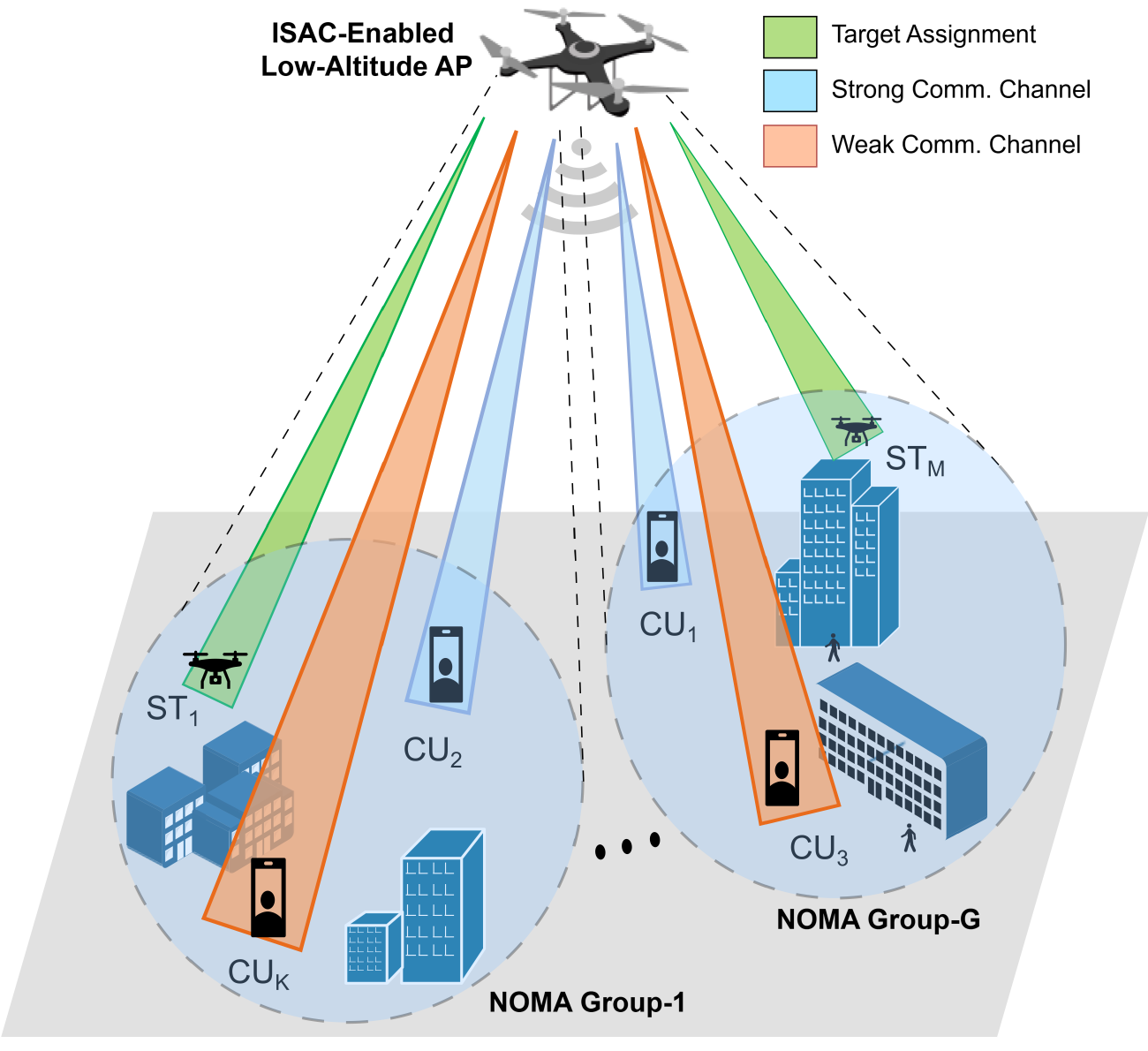}
    \caption{System model of NOMA-Enabled ISAC LAWN.
     }
     \label{fig1}
     \vspace{-8pt}
\end{figure}

\section{System Model}
\label{sec:system_model}
We consider a downlink NOMA-enabled ISAC system in a LAWN. A single ISAC-enabled AP, equipped with an $N_t$ transmit antenna uniform linear array (ULA), serves multiple ground CUs while simultaneously sensing multiple aerial targets as shown in Fig.~\ref{fig1}. The set of CUs is denoted by $\mathcal{K}=\{1,\ldots,K\}$, where $G$ is the number of NOMA groups. The users are divided into a strong user set $\mathcal{U}_s$ and a weak user set $\mathcal{U}_w$ according to their large-scale channel strengths. Each NOMA group consists of a strong user and a weak user. The set of STs is denoted by $\mathcal{M}=\{1,\ldots,M\}$. This work focuses on the one-to-one assignment case with \(M=G\). When \(M>G\), the proposed framework can assign the most suitable subset of STs to the available NOMA groups, while the remaining STs can be scheduled in subsequent sensing frames. When \(M<G\), some NOMA groups may remain unassigned for sensing or may only serve CUs. The extension to general many-to-one or one-to-many group-target assignment is left for future work. The total system bandwidth $B_{\mathrm{tot}}$ is equally divided among the $G$ NOMA groups, and thus each group occupies $B_g=B_{\mathrm{tot}}/G$. Within each group, the strong and weak users share the same group bandwidth using PD-NOMA. The total transmit power is $P_T$, and the power assigned to group $g$ is denoted by $P_g$, with $\sum_{g=1}^{G}P_g\leq P_T$.

\subsection{LAWN Geometry and Channel Model}
The AP is deployed at altitude \(H_{\mathrm{AP}}\). Let \(r_k\) and \(h_k\) denote the horizontal distance and height of CU \(k\), while \(r_m\) and \(h_m\) denote the horizontal distance and altitude of ST \(m\). The corresponding 3D AP-CU and AP-ST distances are given by
\begin{equation}
d_k=\sqrt{r_k^2+\left(H_{\mathrm{AP}}-h_k\right)^2},
\end{equation}
\begin{equation}
d_m=\sqrt{r_m^2+\left(H_{\mathrm{AP}}-h_m\right)^2},
\end{equation}
where \(k\in\mathcal{K}\) and \(m\in\mathcal{M}\). Thus, the AP altitude affects both the communication channel strength and the sensing reliability through the 3D link distances.

Due to the LoS-dominant propagation in LAWNs, the AP-CU channel is modeled using a distance dependent array response as
\newpage
\begin{equation}
\mathbf{h}_k
=
\sqrt{d_k^{-\nu_c}}\,
\mathbf{a}(\widetilde{\theta}_k)e^{j\varphi_k},
\end{equation}
where \(\nu_c\) is the communication path loss exponent, \(\varphi_k\) is the channel phase, and \(\mathbf{a}(\theta)\) is the ULA steering vector defined as
\begin{equation}
\mathbf{a}(\theta)
=
\frac{1}{\sqrt{N_t}}
\left[
1,
e^{j2\pi d_\lambda\sin\theta},
\ldots,
e^{j2\pi d_\lambda(N_t-1)\sin\theta}
\right]^T .
\end{equation}
Here, \(d_\lambda\) denotes the antenna spacing normalized by the carrier wavelength. Although the propagation distances follow the 3D LAWN geometry, the ULA is modeled using a single effective angular direction on the dominant communication-sensing plane; hence, the steering vector captures directional beam alignment, while \(P_g\) and \(d_k^{-\nu_c}\) account for transmit power and path loss. To capture limited mobility and pointing mismatch, we use \(\widetilde{\theta}_k=\theta_k+\Delta\theta_k\) and \(\widetilde{\theta}_m=\theta_m+\Delta\theta_m\), where \(\Delta\theta_k\in[-\Delta_u,\Delta_u]\) and \(\Delta\theta_m\in[-\Delta_t,\Delta_t]\). This quasi-static snapshot model assumes that Doppler shifts and fast trajectory variations are compensated or negligible within one frame, while UPA-based azimuth-elevation beamforming and Doppler-aware modeling are left for future work.

\subsection{NOMA Communication Model}
Let \(s_g\) and \(w_g\) denote the strong and weak users in group \(g\), respectively. The weak user power fraction is \(\alpha_g\in(\alpha_{\min},\alpha_{\max})\), so that the weak and strong user powers are \(\alpha_gP_g\) and \((1-\alpha_g)P_g\), respectively. To support joint communication and sensing, the group beam is constructed by combining the communication beam \(\mathbf{w}_{g}^{\mathrm{c}}=\mathbf{h}_{s_g}/\|\mathbf{h}_{s_g}\|\) and the sensing beam \(\mathbf{w}_{g}^{\mathrm{s}}=\mathbf{a}(\widetilde{\theta}_{\mu(g)})/\|\mathbf{a}(\widetilde{\theta}_{\mu(g)})\|\) as
\begin{equation}
\mathbf{w}_g
=
\frac{
\sqrt{\beta_g}\mathbf{w}_{g}^{\mathrm{c}}
+
\sqrt{1-\beta_g}\mathbf{w}_{g}^{\mathrm{s}}
}{
\left\|
\sqrt{\beta_g}\mathbf{w}_{g}^{\mathrm{c}}
+
\sqrt{1-\beta_g}\mathbf{w}_{g}^{\mathrm{s}}
\right\|
},
\end{equation}
where \(\mu(g)\) is the target assigned to group \(g\), and \(\beta_g\in[0,1]\) is the communication-sensing beam control factor. The transmitted NOMA signal of group \(g\) is
\begin{equation}
\mathbf{x}_g
=
\mathbf{w}_g
\left(
\sqrt{\alpha_gP_g}x_{g,w}
+
\sqrt{(1-\alpha_g)P_g}x_{g,s}
\right),
\end{equation}
where \(x_{g,w}\) and \(x_{g,s}\) are unit power symbols.

The effective channel gains are \(\chi_{g,s}=|\mathbf{h}_{s_g}^{H}\mathbf{w}_g|^2\) and \(\chi_{g,w}=|\mathbf{h}_{w_g}^{H}\mathbf{w}_g|^2\). The weak user SINR, the SINR for decoding the weak user signal at the strong user, and the strong user SINR after SIC are, respectively,
\begin{equation}
\gamma_{g,w}
=
\frac{\alpha_gP_g\chi_{g,w}}
{(1-\alpha_g)P_g\chi_{g,w}+N_0B_g},
\end{equation}
\begin{equation}
\gamma_{g,s\rightarrow w}
=
\frac{\alpha_gP_g\chi_{g,s}}
{(1-\alpha_g)P_g\chi_{g,s}+N_0B_g},
\end{equation}
\begin{equation}
\gamma_{g,s}
=
\frac{(1-\alpha_g)P_g\chi_{g,s}}
{N_0B_g}.
\end{equation}
Accordingly, the weak and strong user rates are
\begin{equation}
R_{g,w}
=
B_g\log_2
\left(
1+\min\{\gamma_{g,w},\gamma_{g,s\rightarrow w}\}
\right),
\end{equation}
\begin{equation}
R_{g,s}
=
B_g\log_2(1+\gamma_{g,s}).
\end{equation}
The system sum rate is \(R_{\mathrm{sum}}=\sum_{g=1}^{G}(R_{g,s}+R_{g,w})\), and the communication QoS constraints require \(R_{g,s}\geq R_s^{\min}\) and \(R_{g,w}\geq R_w^{\min}\) for all groups.

\subsection{Sensing Model}

Each NOMA group is assigned to one ST through the mapping \(\mu(g)\). The sensing gain from group \(g\) to target \(m\) is modeled as
\begin{equation}
\zeta_{g,m}
=
\frac{|\mathbf{a}^{\mathrm H}(\widetilde{\theta}_m)\mathbf{w}_g|^2}{d_m^{\nu_s}},
\label{eq:sensing_gain}
\end{equation}
where \(\nu_s\) is the effective sensing path-loss exponent, \(d_m\) is the AP-target distance, and \(\mathbf{a}(\widetilde{\theta}_m)\) is the steering vector toward target \(m\). The sensing SNR for target assigned to group \(g\) is 
\begin{equation}
\Gamma_{g,\mu(g)}
=
\frac{P_g\zeta_{g,\mu(g)}}{N_0B_g}.
\label{eq:sensing_snr}
\end{equation}

Following the noncoherent energy detection model for a nonfluctuating target~\cite{Detection_Probability}, the detection probability is
\begin{equation}
P_{d,\mu(g)}
=
1-
F_{\chi'^2_{2L}(\lambda_{g,\mu(g)})}(\xi)
=
Q_L
\left(
\sqrt{\lambda_{g,\mu(g)}},
\sqrt{\xi}
\right),
\label{eq:pd_model}
\end{equation}
where \(L=N_{\mathrm{snap}}\), \(\lambda_{g,\mu(g)}=2L\Gamma_{g,\mu(g)}\), and \(\xi=F^{-1}_{\chi^2_{2L}}(1-P_{\mathrm{fa}})\) is the detection threshold for the false-alarm (\(P_{\mathrm{fa}}\)).

The worst-target detection probability and sensing reliability constraint are given by
\begin{equation}
P_d^{\mathrm{worst}}
=
\min_{m\in\mathcal{M}}P_{d,m},
\label{eq:worst_pd}
\end{equation}
\begin{equation}
P_{d,m}\geq P_d^{\mathrm{req}},
\qquad \forall m\in\mathcal{M}.
\label{eq:pd_req}
\end{equation}
The proposed design jointly optimizes user pairing, target assignment, group power, NOMA power split, and beam control factors to improve sum rate while satisfying communication QoS and sensing reliability requirements.

\section{Problem Formulation}
\label{sec:problem_formulation}
The objective is to jointly design the NOMA user pairing, group-target assignment, NOMA group power allocation, and communication-sensing beam control factor for the proposed ISAC-enabled LAWN. Let $\pi$ denote the strong weak user pairing, where $\pi(i)$ is the weak user paired with strong user $i\in\mathcal{U}_s$. Let $\mu(g)$ denote the sensing target assigned to NOMA group $g$. The joint design problem is formulated as
\begin{subequations}
\label{prob:joint_problem}
\begin{align}
\mathcal{P}:
\max_{\pi,\mu,P_g,\alpha_g,\beta_g}
&
R_{\mathrm{sum}}
\label{prob:objective}
\\
\mathrm{s.t.}\quad
& R_{g,s}\geq R_s^{\min},
\qquad \forall g\in\mathcal{G},
\label{prob:strong_qos}
\\
& R_{g,w}\geq R_w^{\min},
\qquad \forall g\in\mathcal{G},
\label{prob:weak_qos}
\\
& P_{d,\mu(g)}\geq P_d^{\mathrm{req}},
\; \forall g\in\mathcal{G},
\label{prob:sensing_qos}
\\
& \textstyle\sum_{g=1}^{G}P_g\leq P_T,
\; P_g\geq 0,\; \forall g\in\mathcal{G},
\label{prob:power_budget}
\\
& \alpha_{\min}\leq \alpha_g\leq \alpha_{\max},
\;\; \forall g\in\mathcal{G},
\label{prob:alpha_constraint}
\\
& 0\leq \beta_g\leq 1,
\qquad \forall g\in\mathcal{G},
\label{prob:beta_constraint}
\\
& \pi(i)\neq \pi(i'),
\; \forall i\neq i',\; \{i,i'\}\in\mathcal{U}_s,
\label{prob:user_pairing}
\\
& \mu(g)\neq \mu(g'), \forall g\neq g',\{g,g'\}\in\mathcal{G}.
\label{prob:target_assignment}
\end{align}
\end{subequations}
Constraints \eqref{prob:strong_qos} and \eqref{prob:weak_qos} guarantee the minimum rate requirements of the strong and weak users, respectively. Constraint \eqref{prob:sensing_qos} ensures that each assigned sensing target satisfies the required detection probability. Constraint \eqref{prob:power_budget} imposes the total transmit-power budget, while \eqref{prob:alpha_constraint} and \eqref{prob:beta_constraint} define the feasible ranges of the NOMA power split and the communication-sensing beam control factor. Finally, \eqref{prob:user_pairing} and \eqref{prob:target_assignment} enforce strong-weak user pairing and group-target assignment, respectively.

\begin{algorithm}[t]
\caption{MUTP Algorithm}
\label{alg:mutp}
\begin{algorithmic}[1]
\REQUIRE Strong user set \(\mathcal{U}_s\), weak user set \(\mathcal{U}_w\), target set \(\mathcal{M}\), channel vectors \(\{\mathbf{h}_k\}\), target steering vectors \(\{\mathbf{a}(\widetilde{\theta}_m)\}\), and AP-target distances \(\{d_m\}\)
\ENSURE User pairing \(\pi^\star\) and group-target assignment \(\mu^\star\)
\STATE Set the initial group power as \(P_g=P_T/G\).
\FOR{each strong user \(i\in\mathcal{U}_s\)}
    \FOR{each weak user \(j\in\mathcal{U}_w\)}
        \STATE Compute \(R_{i,j}^{s}\), \(R_{i,j}^{w}\), and \(R_{i,j}^{\mathrm{sum}}\).
        \STATE Compute the pairing utility \(\Psi_{i,j}\) using \eqref{eq:pair_utility}.
    \ENDFOR
\ENDFOR
\STATE Construct the strong user and weak user preference \\ lists using \(\Psi_{i,j}\).
\STATE Apply stable matching to obtain the strong-weak user pairing \(\pi^\star\).
\STATE Form the NOMA groups according to \(\pi^\star\).
\FOR{each group \(g\in\mathcal{G}\)}
    \FOR{each target \(m\in\mathcal{M}\)}
        \STATE Compute the preliminary sensing suitability \(\zeta_{g,m}^{(0)}\) using \eqref{eq:pre_sensing_gain}.
    \ENDFOR
\ENDFOR
\STATE Select the group-target assignment \(\mu^\star\) using the \\ max-min criterion in \eqref{eq:target_assignment_problem}.
\RETURN \(\pi^\star\) and \(\mu^\star\)
\end{algorithmic}
\end{algorithm}

\section{Matching-Based User-Target Pairing and Adaptive Power Allocation}
\label{sec:proposed_scheme}
The joint problem in \(\mathcal{P}\) involves discrete user pairing and group target assignment variables, as well as continuous power allocation and beam control factors. Direct exhaustive search over all possible associations is computationally expensive, especially as the number of CUs and STs increases. To address this challenge, we develop a low-complexity MUTP-APA framework that first constructs communication and sensing-aware associations and then refines the power allocation through QoS and sensing-aware APA.

\subsection{Matching-Based User-Target Pairing}
\label{subsec:mutp}
The first module forms NOMA groups and assigns STs to the resulting groups. Let \(i\in\mathcal{U}_s\) denote a strong user and \(j\in\mathcal{U}_w\) denote a weak user. For each candidate pair \((i,j)\), the achievable NOMA rates are evaluated under an initial equal group power allocation \(P_g=P_T/G\). At this preliminary pairing stage, the rates are computed using the communication-oriented beam \(\mathbf{w}_{i}^{(0)}=\mathbf{h}_{i}/\|\mathbf{h}_{i}\|\), a fixed initial power split \(\alpha_0\), and \(\beta_g=1\). Define the QoS-feasible event as
\begin{equation}
\mathcal{F}_{i,j}
\triangleq
\left\{
R_{i,j}^{s}\geq R_s^{\min},
\;
R_{i,j}^{w}\geq R_w^{\min}
\right\},
\label{eq:qos_feasible_pair}
\end{equation}
and the normalized QoS margin as
\begin{equation}
\eta_{i,j}
=
\min
\left\{
\frac{R_{i,j}^{s}}{R_s^{\min}},
\frac{R_{i,j}^{w}}{R_w^{\min}}
\right\}.
\label{eq:qos_margin_pair}
\end{equation}
The pair utility is defined as
\begin{equation}
\Psi_{i,j}
=
\begin{cases}
R_{i,j}^{\mathrm{sum}}+\epsilon_q\eta_{i,j},
& \mathcal{F}_{i,j},\\
\epsilon_o R_{i,j}^{\mathrm{sum}},
& \mathrm{otherwise},
\end{cases}
\label{eq:pair_utility}
\end{equation}
where \(R_{i,j}^{\mathrm{sum}}=R_{i,j}^{s}+R_{i,j}^{w}\), \(\epsilon_q\) is a small QoS-margin weight, and \(\epsilon_o\ll1\) penalizes infeasible pairings. Based on \(\Psi_{i,j}\), stable matching is used to form the  NOMA groups.

After user pairing, each NOMA group is assigned to a sensing target. Since target assignment is performed before the final adaptive beam selection, a preliminary communication beam \(\mathbf{w}_g^{\mathrm{c}}\) is used to evaluate the group-target sensing suitability as
\begin{equation}
\zeta_{g,m}^{(0)}
=
\frac{
\left|\mathbf{a}^{\mathrm H}(\widetilde{\theta}_m)
\mathbf{w}_{g}^{\mathrm{c}}\right|^2
}{
d_m^{\nu_s}
}.
\label{eq:pre_sensing_gain}
\end{equation}
The group-target assignment is obtained by maximizing the weakest assigned sensing gain:
\begin{equation}
\begin{aligned}
\mu^\star
&=
\arg\max_{\mu}
\left[
\min_{g\in\mathcal{G}}
\zeta_{g,\mu(g)}^{(0)}
\right],
\\
&\quad
\mathrm{s.t.}\quad
\mu(g)\neq \mu(g'),\ \forall\, g\neq g'.
\end{aligned}
\label{eq:target_assignment_problem}
\end{equation}
If more than one assignment gives the same minimum sensing gain, the assignment with the larger total sensing gain is selected. Since this work considers \(M=G\) with a small number of STs, (22) is solved by enumerating all \(M!\) possible one-to-one group-target assignments and selecting the assignment with the largest minimum sensing gain. The MUTP procedure is summarized in Algorithm~\ref{alg:mutp}. The algorithm first evaluates the QoS-aware pairing utility for all possible strong-weak user pairs and constructs preference lists for stable matching. After forming the NOMA groups, it evaluates the preliminary sensing suitability between each group and each target and then selects the group-target assignment that maximizes the weakest assigned sensing gain. In this way, Algorithm~\ref{alg:mutp} jointly accounts for communication feasibility and sensing reliability while avoiding exhaustive association search.

\subsection{QoS and Sensing-Aware Adaptive Power Allocation}
\label{subsec:apa}
Given \(\pi^\star\) and \(\mu^\star\), the APA stage selects the group powers \(P_g\), NOMA power fractions \(\alpha_g\), and beam-control factors \(\beta_g\). Instead of solving the continuous nonconvex problem, APA evaluates a finite set of normalized candidate power profiles, including uniform, sensing-aware, QoS-aware, waterfilling-inspired communication-oriented, and hybrid profiles. Each profile is generated as
$P_g^{(c)}=\frac{P_T\omega_g^{(c)}}{\sum_{\ell=1}^{G}\omega_\ell^{(c)}},$
where \(\omega_g^{\mathrm{U}}=1\), \(\omega_g^{\mathrm{S}}=1/(\zeta_{g,\mu(g)}+\epsilon)\), \(\omega_g^{\mathrm{C}}=\max\{\chi_{g,s},\chi_{g,w}\}\), and \(\omega_g^{\mathrm{H}}\) is a weighted sensing-communication combination. The QoS-aware profile is obtained from the estimated minimum group power required to satisfy \(R_s^{\min}\) and \(R_w^{\min}\). The NOMA power splitting and beam control candidates are selected from
\(\mathcal{A}=\{\alpha_{\min},\alpha_{\min}+\Delta_{\alpha},\ldots,\alpha_{\max}\}\)
and
\(\mathcal{B}=\{\beta_1,\ldots,\beta_{N_{\beta}}\}\subset[0,1]\), respectively.

For each candidate power profile, the NOMA power fraction is selected from the feasible set \(\mathcal{A}\), and the beam-control factor is selected from \(\mathcal{B}\). The resulting sum rate, worst-target detection probability, average detection probability, communication QoS status, and sensing reliability status are then evaluated. If at least one candidate satisfies the QoS and sensing constraints, the final APA solution \(\{P_g^\star,\alpha_g^\star,\beta_g^\star\}\) is selected as the candidate with the highest sum rate, otherwise, the candidate with the best joint communication-sensing performance is selected. Since APA searches over a finite candidate set, it provides a low-complexity candidate-based solution rather than a globally optimal solution of \(\mathcal{P}\). The performance gap depends on the richness of the candidate power profiles and the resolutions of \(\mathcal{A}\) and \(\mathcal{B}\), increasing these sets can reduce the gap at the cost of higher complexity. The complexity of the proposed NISAC-MUTP-APA scheme is \(\mathcal{O}(G^2+GM+M!G+N_PN_{\alpha}N_{\beta}G)\), where \(G=K/2\), \(M\) is the number of STs, and \(N_P\), \(N_{\alpha}\), and \(N_{\beta}\) denote the numbers of candidate power profiles, NOMA power splitting candidates, and beam control candidates, respectively. In contrast, the exhaustive search benchmark enumerates all strong-weak user pairing and group target assignment combinations, leading to complexity on the order of \(\mathcal{O}(G!M!+G!M!N_PN_{\alpha}N_{\beta}G)\).

\section{Simulation Results and Discussion}
\label{sec:results}
This section evaluates the performance of the proposed NISAC-MUTP scheme and compares it with fixed GP, exhaustive search pairing, and orthogonal ISAC (OISAC) benchmark schemes under fixed power allocation (FPA) and APA. The main simulation parameters are summarized in Table~\ref{tab:sim_params}.

\begin{figure}[t]
    \centering
    \includegraphics[width=0.85\linewidth]{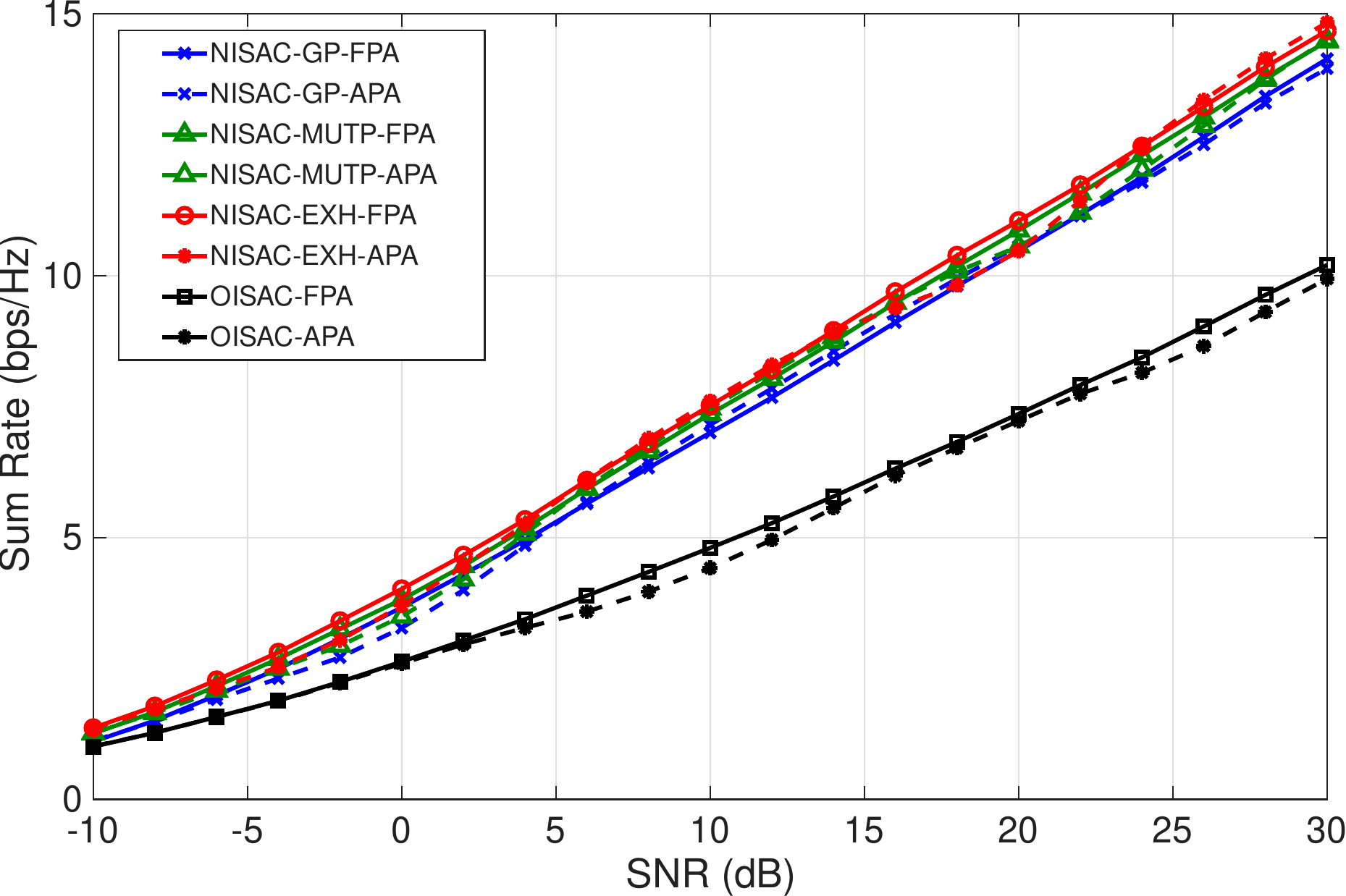}
    \caption{Sum rate comparison of the proposed MUTP and other schemes.}
    \vspace{-8pt}
    \label{fig:sum_rate_snr}
\end{figure}

\begin{table}[t]
\centering
\caption{Simulation parameters used in the paper.}
\label{tab:sim_params}
\renewcommand{\arraystretch}{1.12}
\begin{tabular}{|p{0.52\linewidth}|p{0.32\linewidth}|}
\hline
\textbf{Parameter} & \textbf{Value} \\
\hline
Number of transmit antennas ($N_t$) & 16 \\
\hline
Number of CUs ($K$) & 10 \\
\hline
Number of STs ($M$) & 5 \\
\hline
Normalized total bandwidth ($B_{\mathrm{tot}}$) & 1  \\
\hline
Normalized transmit power ($P_T$) & 1  \\
\hline
Number of Monte Carlo trials & $10^5$ \\
\hline
AP altitude ($H_{\mathrm{AP}}$) & 100 m \\
\hline
Target height range & 30-120 m \\
\hline
False-alarm probability ($P_{\mathrm{fa}}$) & $10^{-3}$ \\
\hline
Required detection probability ($P_d^{\mathrm{req}}$) & 0.7 \\
\hline
Sensing snapshots ($N_{snap}$) & 100 \\
\hline
\end{tabular}
\end{table}

\begin{figure}[t]
    \centering
    \includegraphics[width=0.85\linewidth]{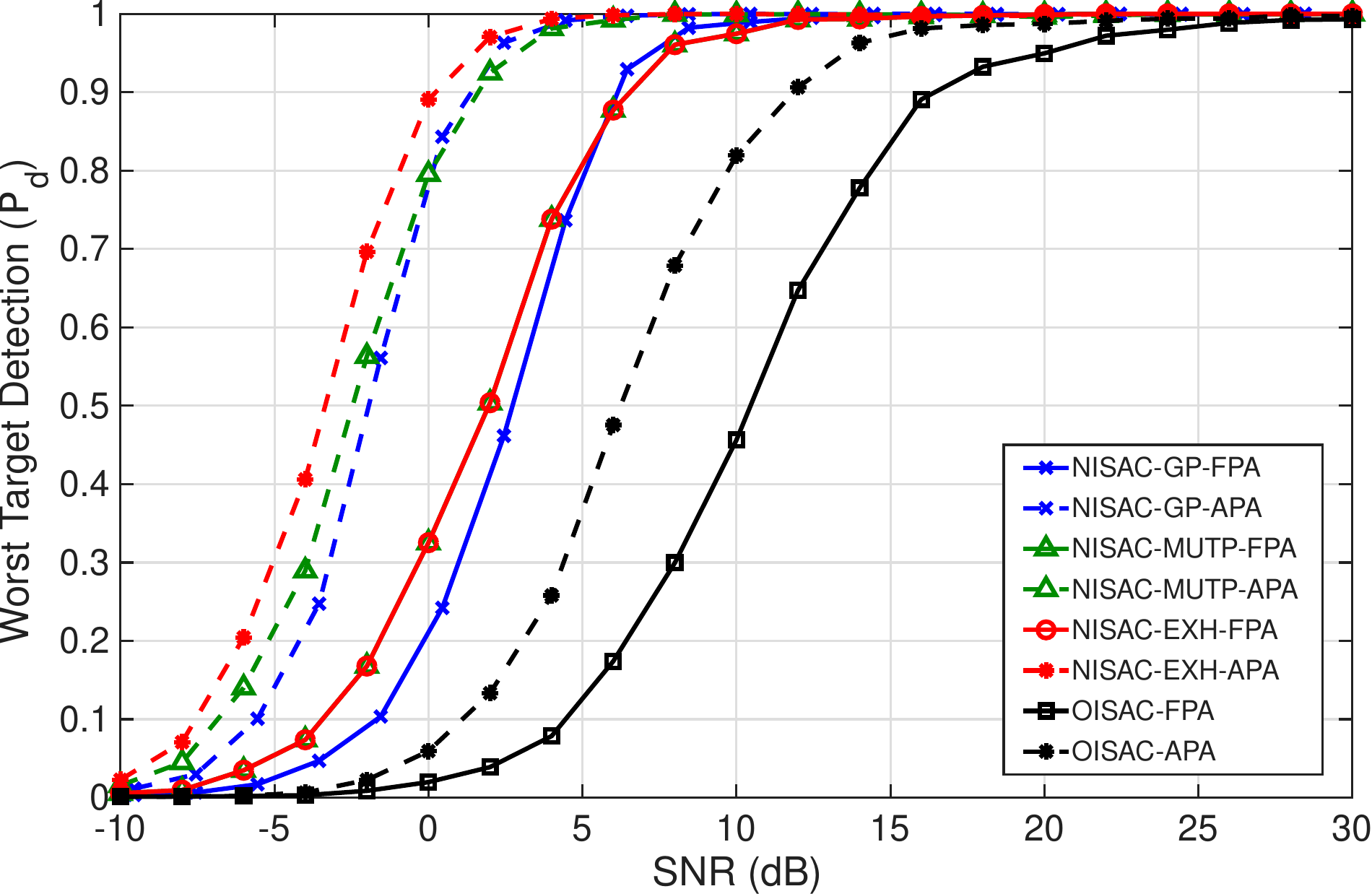}
    \caption{Worst-target detection probability comparison of the schemes.}
    \vspace{-8pt}
    \label{fig:wtd_snr}
\end{figure}

Fig.~\ref{fig:sum_rate_snr} shows the sum rate performance at different SNR values. The OISAC benchmarks achieve the lowest sum rate because orthogonal access divides the available group resources between users. In contrast, the NISAC-based schemes allow the paired users to share the same group bandwidth through PD-NOMA, which improves spectral efficiency. Among the NISAC schemes, the exhaustive search benchmark provides the highest performance, while the proposed NISAC-MUTP scheme closely approaches it with lower pairing complexity. This confirms that the proposed scheme can achieve near-exhaustive performance without requiring exhaustive search.

Fig.~\ref{fig:wtd_snr} presents the worst-target detection probability versus SNR. The worst-target metric is important because it reflects the reliability of the weakest sensed target. The proposed NISAC-MUTP-APA scheme reaches the required detection probability at a lower SNR compared with the fixed GP and OISAC benchmarks. This improvement is due to the joint effect of sensing-aware target assignment and APA. The exhaustive search scheme provides the best benchmark performance, while the proposed MUTP scheme follows it closely, demonstrating the effectiveness of the proposed low-complexity pairing design.

\begin{figure}[t]
    \centering
    \subfloat[Communication outage probability (COP) of all schemes.]
    {
        \includegraphics[width=0.85\linewidth]{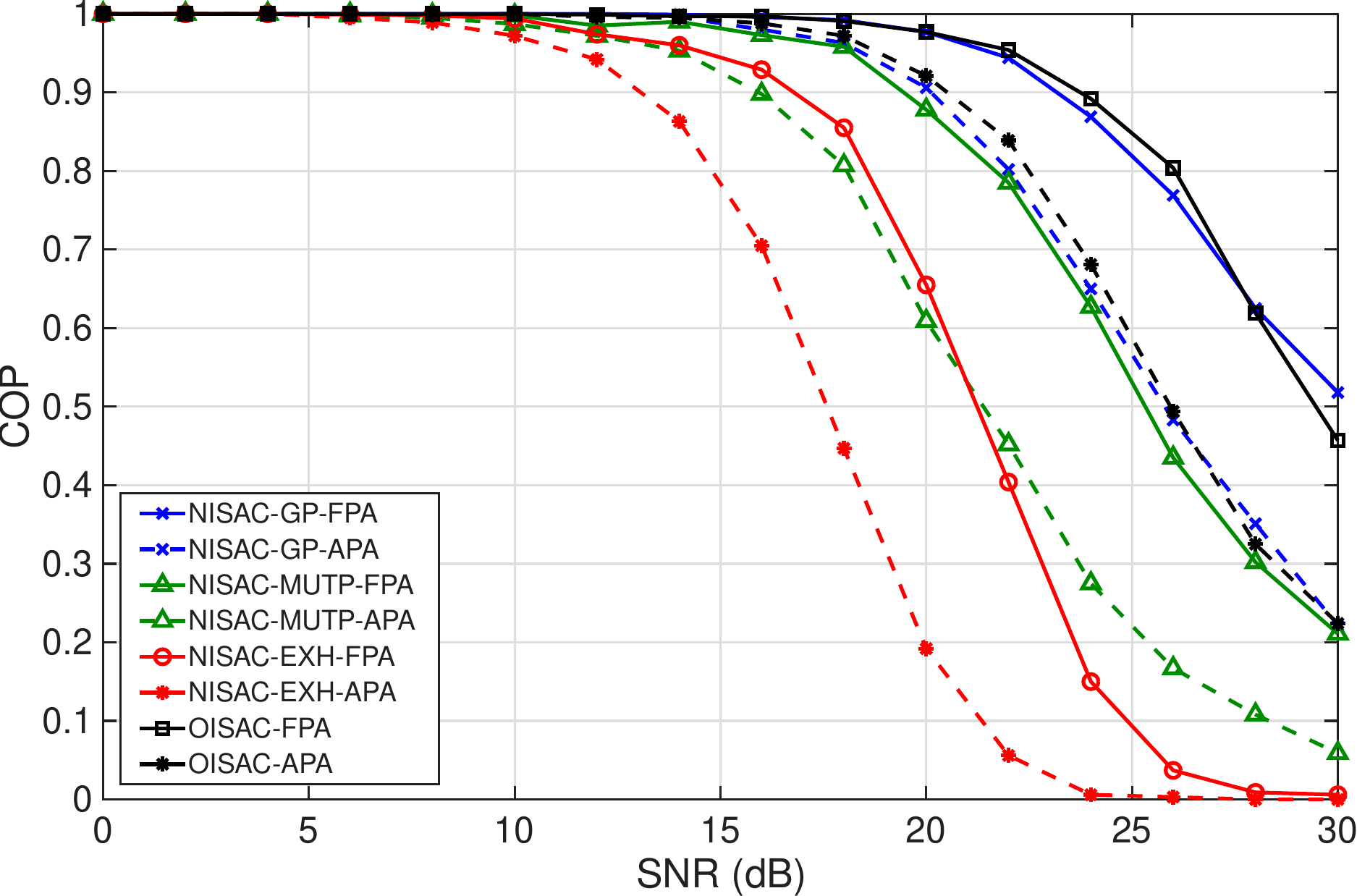}
        \label{fig:cop_snr}
    }
    \hfill
    \subfloat[Sensing outage probability (SOP) of the proposed and benchmark schemes.]
    {
        \includegraphics[width=0.85\linewidth]{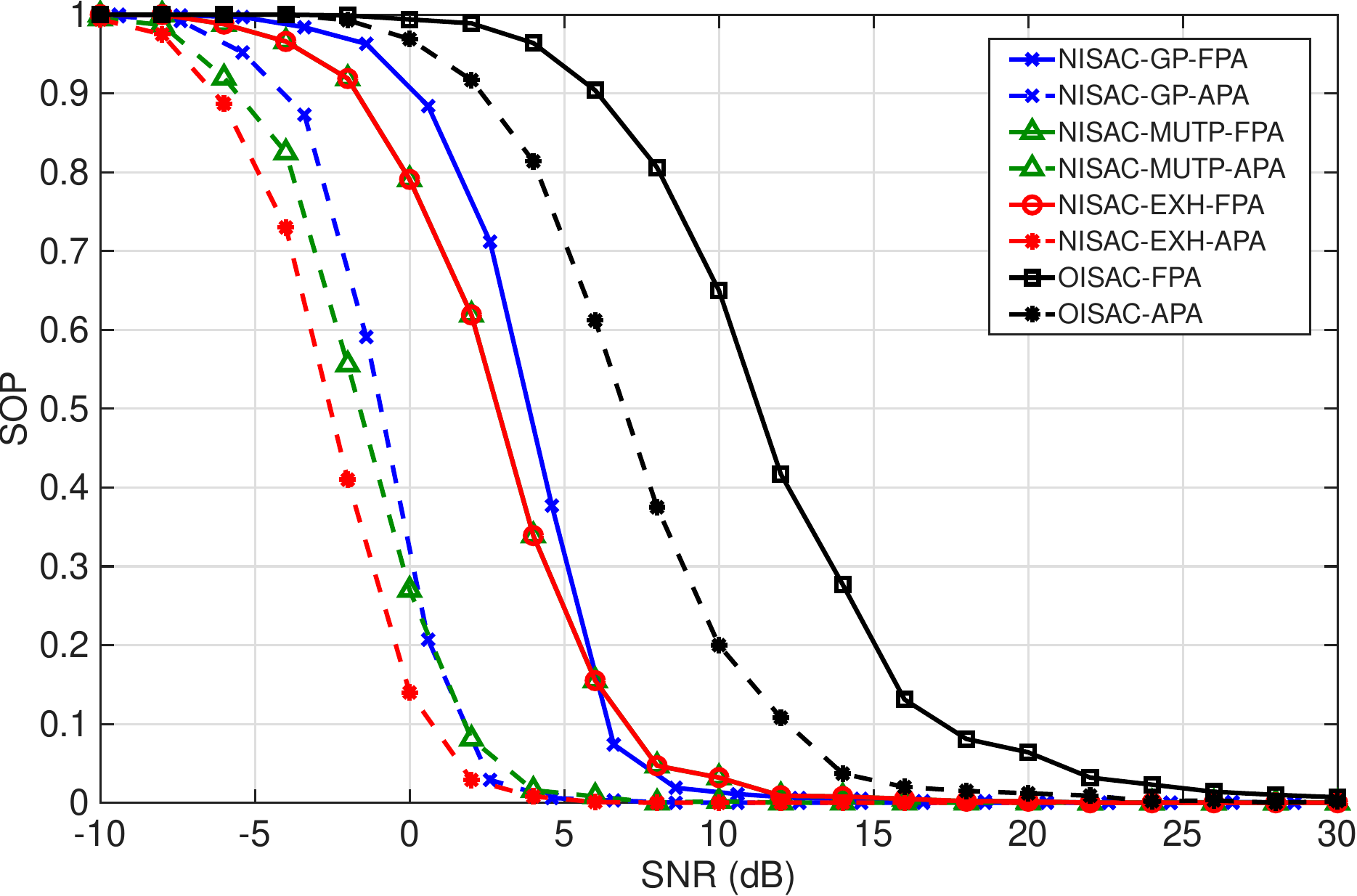}
        \label{fig:sop_snr}
    }
    \caption{Outage probability comparison of all schemes.}
    \vspace{-8pt}
    \label{fig:outage_snr}
\end{figure}

In Fig.~\ref{fig:cop_snr}, the proposed NISAC-MUTP-APA scheme significantly reduces the probability of violating the user QoS constraints compared with the fixed GP and OISAC benchmarks. The COP reduction is more visible in the medium to high SNR region, where proper user pairing and APA become more effective. It is observed from Fig.~\ref{fig:sop_snr} that the SOP of the proposed NISAC-MUTP-APA scheme decreases rapidly as the SNR increases. This is because SOP is determined by the event that at least one assigned target fails to satisfy the required detection probability. The proposed sensing-aware target assignment improves the weakest group-target sensing links, while APA allocates resources more effectively to support detection reliability. As a result, the probability of violating the sensing constraint is significantly reduced. The OISAC benchmark requires higher SNR to reduce sensing outage because orthogonal bandwidth allocation limits the effective sensing opportunity. In contrast, the NISAC schemes exploit the shared communication-sensing waveform more efficiently.

\section{Conclusion}
\label{sec:conclusion}
This paper proposed a NOMA-enabled ISAC scheme for LAWNs, where ground CUs and aerial STs are jointly considered in an altitude-aware system. An MUTP scheme was developed to form strong-weak NOMA user pairs and assign STs to NOMA groups, while an APA strategy was introduced to balance sum rate, user QoS, and sensing reliability. Simulation results showed that the proposed NISAC-MUTP-APA scheme improves the sum rate, worst-target detection probability, communication and sensing outage compared with GP and OISAC benchmarks, while closely approaching exhaustive-search performance with lower association complexity. Future work will consider UPA-based 3D beamforming, Doppler-aware mobility modeling, imperfect CSI, and UAV trajectory optimization.

\end{document}